\documentclass[aps,prd,superscriptaddress,twocolumn]{revtex4-2}
\usepackage[utf8]{inputenc}
\usepackage{dsfont}
\usepackage{tensor}
\usepackage{todonotes}
\usepackage{hyperref}
\hypersetup{
    colorlinks,
    citecolor=navyblue,
    filecolor=navyblue,
    linkcolor=navyblue,
    urlcolor=navyblue
}
\usepackage{amsthm}
\usepackage{amsmath}
\usepackage[noabbrev]{cleveref}
\usepackage{amsfonts}
\usepackage{comment}
\allowdisplaybreaks[4]        
\usepackage{amssymb}
\usepackage{euscript}     
\usepackage{braket}
\usepackage{starfont}
\usepackage{color,soul}         
\usepackage{tensor}        

\usepackage{lipsum} 
\usepackage{graphicx}
\usepackage{slashed}
\usepackage{leftidx}
\usepackage{subfigure}
\usepackage{bbm}
\usepackage{bm}
\definecolor{outerspace}{rgb}{0.25, 0.29, 0.3}
\definecolor{scarlet}{rgb}{1.0, 0.13, 0.0}
\usepackage[header,title,page,titletoc]{appendix}  
\definecolor{princetonorange}{rgb}{1.0, 0.56, 0.0}
\definecolor{WildStrawberry}{rgb}{1.0, 0.26, 0.64}
\definecolor{rossocorsa}{rgb}{0.83, 0.0, 0.0}
\definecolor{navyblue}{rgb}{0.0, 0.0, 0.5}
\usepackage{float}
\usepackage[paper=letterpaper,margin=1in]{geometry}

\usepackage{mathtools}

\DeclareMathAlphabet{\pazocal}{OMS}{zplm}{m}{n}

\newcommand{\bea}{\begin{eqnarray}}

\newcommand{\eea}{\end{eqnarray}}
\newcommand{\ba}{\begin{eqnarray}}
\newcommand{\ea}{\end{eqnarray}}

\newcommand{\be}{\begin{equation}}
\newcommand{\ee}{\end{equation} }

\newcommand{\beqa}{\begin{eqnarray}}
\newcommand{\eeqa}{\end{eqnarray}}
\newcommand{\beqar}{\begin{eqnarray*}}
\newcommand{\eeqar}{\end{eqnarray*}}

\DeclareMathOperator{\tr}{tr}

\def\({\left (}
\def\){\right )}

\makeatletter
\newcommand{\dal}{\mathop{\mathpalette\dal@\relax}}
\newcommand{\dal@}[2]{%
  \begingroup
  \sbox\z@{$\m@th#1\square$}%
  \dimen0=\fontdimen8
    \ifx#1\displaystyle\textfont\else
    \ifx#1\textstyle\textfont\else
    \ifx#1\scriptstyle\scriptfont\else
    \scriptscriptfont\fi\fi\fi3
  \makebox[\wd\z@]{%
    \hbox to \ht\z@{%
      \vrule width \dimen0
      \kern-\dimen0
      \vbox to \ht\z@{
        \hrule height \dimen0 width \ht\z@
        \vss
        \hrule height 2\dimen0
      }%
      \kern-2.5\dimen0
      \vrule width 2.5\dimen0
    }%
  }%
  \endgroup
}
\makeatother

\begin{document}

\title{Mutual Information from Modular Flow in General CFTs}
\author{C\'esar A. Ag\'on}
\email{cesar.agon@upct.es}
\affiliation{Departamento de F\'isica Aplicada y Tecnolog\'ia Naval, Universidad Polit\'ecnica de Cartagena member of European University of Technology EUT+, Campus Muralla del Mar, C/Dr Fleming S/N, 30202 Cartagena, Spain}

\author{Pablo Bueno}
\email{pablobueno@ub.edu}
\affiliation{Departament de F\'isica Qu\`antica i Astrof\'isica, Institut de Ci\`encies del Cosmos\\
 Universitat de Barcelona, Mart\'i i Franqu\`es 1, E-08028 Barcelona, Spain }

\author{Adem Deniz Piskin}
\email{a.d.piskin@tudelft.nl}
\affiliation{Institute for Theoretical Physics, Utrecht University, 3584 CC Utrecht, The Netherlands}
\affiliation{QuTech, Delft University of Technology, Lorentzweg 1, 2628 CJ Delft, The Netherlands}

\author{Guido van der Velde}
\email{guidovandervelde@icc.ub.edu}
\affiliation{Departament de F\'isica Qu\`antica i Astrof\'isica, Institut de Ci\`encies del Cosmos\\
 Universitat de Barcelona, Mart\'i i Franqu\`es 1, E-08028 Barcelona, Spain }


\begin{abstract}
The vacuum mutual information (MI) of subregion algebras provides a universal window into the data of general conformal field theories (CFTs). Exploiting the geometric nature of the modular flow associated to ball-shaped regions and the operator product expansion of twist operators implementing the replica symmetry in an $n$-fold version of a CFT, it is possible to construct a hierarchy of increasingly refined approximations to the full MI.  In this letter, we use the two-point functions of primaries of arbitrary spin in the replicated theory to constrain the twist operators, and find their contribution to the MI of arbitrarily boosted balls in any $d$-dimensional CFT. When the two-point functions involve the primary with the lowest scaling dimension, our result provides the most precise approximation for the long-distance behavior of the MI, superseding all previous expansions. Building upon this result and certain universal properties of the short- and long-distance regimes, we put forward a new high-precision analytic approximation to the MI for arbitrary separations. The accuracy of our approach is validated against exact $d=2$ and lattice $d=3$ results. We further apply it to characterize the MI of a $d=4$ Maxwell field, a case for which no prior results are available. \\
\begin{center}\textit{Dedicated to the memory of Umut G\"ursoy}
\end{center}
\end{abstract}
\maketitle

 The vacuum mutual information (MI) of operator algebras associated to spacetime regions is expected to fully characterize any given QFT. On the one hand, the MI provides a universal geometric regulator for entanglement entropy, capturing all the associated universal coefficients ~\cite{Casini:2007dk,Casini:2008wt,Casini:2014yca,Casini:2015woa,Bueno:2021fxb,Huerta:2022cqw,Bueno:2023gey,Bueno:2026hku}.
Complementarily, it has been argued that a long-distance expansion of the MI should allow for the systematic extraction of the full underlying QFT data ~\cite{Headrick:2010zt,Calabrese:2010he,Long:2016vkg,Chen:2016mya,Chen:2017hbk,Cardy:2013nua,Agon:2015twa,Agon:2015ftl,Agon:2021zvp,Casini:2021raa,Agon:2021lus,Agon:2022efa,Agon:2024zae,Agon:2024xvs}.

The second approach relies on two key facts: i) the geometric nature of the modular flow associated with ball regions in any conformal field theory (CFT) \cite{Casini:2011kv,Hislop:1981uh}; and ii) the possibility of expressing the MI in terms of a correlator of two \emph{twist operators},  which implement the ${\mathbb Z}_n$ orbifolding of the $n$ copies of the theory CFT$^{\otimes n}$ \cite{Calabrese:2004eu,Calabrese:2005zw,Hung:2014npa,Casini:2005rm}. These features allow for increasingly sophisticated and complete approximations to the MI. These arise from imposing certain consistency conditions that connect the expectation value of the twist operators probed by local operators in different copies of the CFT to the expectation value of modularly-evolved operators in the original theory \cite{Casini:2021raa,Agon:2022efa,Agon:2024xvs}.

\begin{figure*}[t!]
\centering 
\includegraphics[width=1\textwidth]{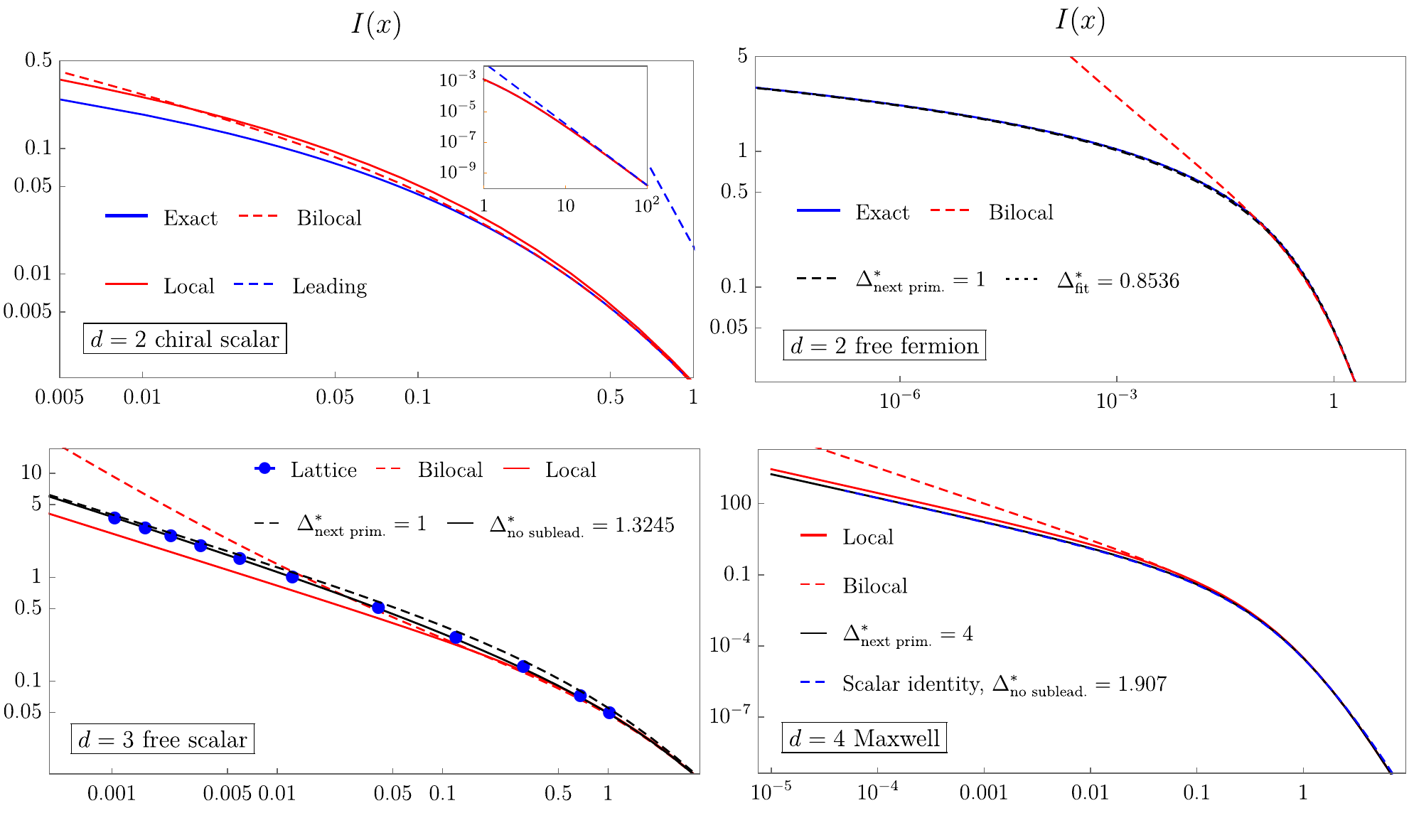}
\caption{
Approximations to the MI of pairs of parallel disjoint balls for various free theories. In all cases, $x\equiv D/L$, where $D$ is the distance between the regions' closest boundaries, and $L$ is the length of both intervals in $d=2$, and the disks/spheres radii in $d=3,4$. Both for the free chiral scalar and fermion in $d=2$ (upper row) the exact results are known  \cite{Casini:2005rm,Casini:2009vk,Arias:2018tmw} and they are plotted in blue. Numerical results from the lattice---with conservative error bars corresponding to the size of the dots---are presented for the $d=3$ free scalar. In all plots, the two red curves represent approximations arising from the modular flow of the two-copy sector, both of which become exact at long distances. While the bi-local ansatz (dashed) is consistently more accurate than the local one (solid) at intermediate separations---with the exception of the $d=2$ free fermion, for which the latter is identical to the exact result \cite{Casini:2008wt,Agon:2021zvp}---, it goes astray at short distances, due to its volume-law behavior.
The black curves represent refinements of the bi-local approximation in which the number of spacetime dimensions is switched, $d\rightarrow d-1$, and an additional contribution from an operator of dimension $\Delta^*$ and $\sigma$ degrees of freedom is added so that the correct area-law is recovered without spoiling the long-distance behavior---see Eqs.\,(\ref{MI_short_distance_ansatz}) and (\ref{sigm}). In the plots, $\Delta^*_{\rm fit}$ corresponds to a least-squares fit to the exact result, $\Delta^*_{\rm next\,prim.}$ to adding the next-to-leading primary contribution and $\Delta^*_{\rm no\,sublead.}$ to the choice which removes the spurious subleading term in the short-distance expansion.
}
\label{fig:Mutual}
\end{figure*}

Using this approach, we present the contribution to the MI of arbitrarily boosted balls arising from the two-copy sector associated to a primary operator of arbitrary spin in a general $d$-dimensional CFT. Our formula supersedes and generalizes all previous approximations to the MI long-distance expansion. Additionally, we use this formula to put forward a new analytic high-precision approximation to the MI of any CFT valid for arbitrary separations. We show that this accurately reproduces the exact MI curve for a $d=2$ free fermion and  lattice results for a $d=3$ free scalar, and we use it to predict the result for a $d=4$ Maxwell field---see Fig.\,\ref{fig:Mutual}.

{\bf Mutual information and modular flows:} The MI between two disjoint spacetime regions $A,B$, can be defined in terms of the von Neumann entropies of the corresponding reduced density matrices, $ I(A,B)\equiv S_A+S_B-S_{A\cup B}$. Here, $S_A\equiv -{\rm Tr} [\rho_A\log \rho_A]$, where $\rho_A\equiv {\rm Tr}_{\bar A} \ket{\Omega}\bra{\Omega}$, and from now on we will consider the vacuum $\ket{\Omega}$ as our global state. While all the individual entropies appearing in the above formula suffer from UV divergences, they always cancel one another in $I(A,B)$, which is a well-defined quantity in the continuum \cite{Witten:2018zxz,Casini:2022rlv}. 

A standard technique for evaluating the MI for a given CFT entails evaluating the expectation value of two properly normalized \emph{twist operators}  defined in CFT$^{\otimes n}$ \cite{Calabrese:2004eu,Calabrese:2005zw,Hung:2014npa}.
The precise formula reads
 \begin{align}\label{mutual-twists}
    &I(A,B) = \lim_{n\rightarrow1}\frac{1}{n-1} \langle \Tilde{\Sigma}_A^{(n)} \Tilde{\Sigma}_B^{(n)}\rangle\,.
\end{align}
 Here $\Sigma^{(n)}_X\equiv\langle \Sigma^{(n)}_X\rangle( {\mathbf{1}}+\tilde{\Sigma}^{(n)}_X)$ and $\Sigma^{(n)}_X$ implements the global cyclic symmetry transformation to operators localized in region $X$, namely $\Sigma^{(n)}_X :{\cal O}^i_X\to {\cal O}^{i+1}_X$, while acting trivially on operators outside $X$ \cite{Calabrese:2004eu,Hung:2014npa}.  The index $i$ labels the copies, subject to the cyclic identification $n+1\equiv 1$. 

 In order to characterize $\tilde \Sigma_A^{(n)}$, we may consider  its correlation function with an arbitrary set of operators. Here we will be interested in contributions arising from operators supported on two distinct copies, $1\leq l \neq k\leq n$. As a consequence, it will suffice to invoke the following result, which holds in the $n\rightarrow 1$ limit \cite{Casini:2021raa,Agon:2022efa,Agon:2024xvs},\footnote{Note that this definition is in agreement with the KMS condition, which leads to the relation $G_{\alpha\beta}(x_1,x_2,-s)=(-1)^F G_{\alpha\beta}(x_1,x_2,s)$, where $G_{\alpha\beta}(x_1,x_2,s)$ is the left hand side of Eq. (\ref{eq:two-point-twist}), and $s\equiv i\tau_{lk}$. Moreover, $F=1$ for fermions while $F=0$ for bosons. Since $s\rightarrow -s$ corresponds to the exchange $l\rightarrow k$, this reflects the symmetric (antisymetric) nature of bosonic (fermionic) fields. See Appendix B of \cite{Agon:2024zae}. }
      \begin{align}\notag
&\langle\tilde{\Sigma}_A^{(n)}{\cal O}_\alpha^l(x_1){\cal O}_\beta^k(x_2)\rangle \overset{\small n\rightarrow 1}{=} \\ &\! \quad\begin{cases}
 \langle {\cal O}_\alpha (x_1) {\cal O}^A_{\beta}[x_2,i\tau_{lk}]\rangle \qquad (l>k)\\
      \langle {\cal O}^A_{\beta}[x_2,i\tau_{kl}]{\cal O}_\alpha (x_1)\rangle \qquad (l<k)
      \end{cases} \label{eq:two-point-twist} 
\end{align}
This remarkable formula requires some explanation. On the one hand, $\mathcal{O}_\alpha^l(x)$ is a local operator, with a possible spin index $\alpha$, defined in the $l$-th copy  of the CFT at the spacetime point $x$. Also, we defined ${\cal O}^A_\alpha[x,s] \equiv  \rho_A^{-is} {\cal O}_\alpha(x)\rho_A^{is}$, which is the operator resulting from the modular-flow evolution associated to $\rho_A$, and we introduced the notation $\tau_{kl}\equiv \pi(k-l)/n$. The correlators in the right hand side are analytic and well-defined in the region $\text{Im}(s)\in(0,1)$, which is the case for $s=i|\tau_{lk}|$.  
Hence,  Eq.\,\eqref{eq:two-point-twist} connects a correlation function in the replicated theory with a correlation function of modularly transformed operators in the seed CFT. 

 For spherical entangling surfaces, the  transformation that implements the modular flow corresponds to a conformal transformation 
\cite{Casini:2011kv,Hislop:1981uh},
\begin{align}
    x^\pm_A[\xi,s] = R_A\frac{(R_A+\xi^\pm)-e^{\mp2\pi s}(R_A-\xi^\pm)}{(R_A+\xi^\pm)+e^{\mp 2\pi s}(R_A-\xi^\pm)}\,,\label{mod-flow}
\end{align}
where $R_A$ is the radius of the sphere $A$ and $\xi^\pm=r\pm t$ are null coordinates with respect to its center.  In that case, the relevant correlator 
takes the form
\bea\label{corr-modular}
&&\!\!\!\! \!\!\!\!\!\!\langle {\cal O}_\alpha (x_1) {\cal O}^A_{\beta}[x_2,s]\rangle\\
&&\quad \!\! \!\!\!\!=\Big|\frac{\partial x_{2,A}[s]}{\partial x_2}\Big|^\Delta\!\! {\mathcal R}_{\beta}^{\,\,\,\lambda}(\Lambda_A(x_2,s)) \, G_{\alpha \lambda}(x_1, x_{2,A}[s])\,, \nonumber
\eea 
where we introduced the notation $x_{2,A}[s]\equiv x_A[x_2,s]$. Also, ${\mathcal R}_{\beta}^{\,\lambda}(\Lambda_A(x_2,s))$ is the local Lorentz transformation associated to the conformal transformation \eqref{mod-flow} based at $x_2$ in the representation ${\mathcal R}$ of $\mathcal{O}$ and $|{\partial x_{2,A}[s]}/{\partial x_2}|$ is its associated scaling factor. Finally, $G_{\alpha\lambda}$ denotes the 
two-point function of $\mathcal{O}$, namely,
$G_{\alpha \beta}(x,y) \equiv \langle {\cal O}_\alpha(x){\cal O}_\beta(y)\rangle =H_{\alpha\beta}(x-y)\cdot|x-y|^{-2\Delta}$,
 where $H_{\alpha\beta}(x)$ captures the tensor structure of the correlator and $\Delta$ is the scaling dimension of ${\cal O}$.

Our goal is to evaluate the contribution to the MI arising from the part of the twist operator which is responsible for making \eqref{eq:two-point-twist} hold.
Analogous formulas for higher-point correlators can be derived following \cite{Agon:2024xvs}, which would allow one to capture multiple-copy contributions to the MI. 
 
An arbitrary operator in the replica theory can be written as a linear combination of products of operators associated to the individual copies. 
The most general operator that can contribute to the left hand side of \eqref{eq:two-point-twist} can be written as a double integral with support in the causal completion of the entangling region that is generated from the product of two primaries in two different replicas $l\neq k$, namely,
\begin{align}
    &\Tilde{\Sigma}_A^{(n)} = \sum_{l\neq k}\iint_{D_A}\mathcal{G}_{A,\alpha\beta}^{lk}(\xi_1,\xi_2)O^l_\alpha(\xi_1)O^k_\beta(\xi_2)\,,
\label{eq:bilocaldef}
\end{align}
where we denoted $\iint_{D_A}\equiv \int_{D_A}\int_{D_A} {\rm d}^d\xi_1 {\rm d}^d\xi_2$. 
Plugging ansatz \eqref{eq:bilocaldef} into \eqref{eq:two-point-twist} leads to 
\begin{align}
&\langle\tilde{\Sigma}_A^{(n)}O_\alpha^l(x_1)O_\beta^k(x_2)\rangle\!\!\overset{\small n\rightarrow 1}{=} \label{Twist-action-corrs}
 \\
&\quad 2 \iint_{D_A}\,\mathcal{G}_{A,\alpha\beta}^{lk}(\xi_1,\xi_2) G_{\alpha \mu}(\xi_1, x_1)G_{\beta \nu}(\xi_2, x_2)\,, \nonumber
\end{align}
which can be inverted to find the kernel function $\mathcal{G}_{A,\alpha\beta}^{lk}(\xi_1,\xi_2)$. 
We define the inverse correlator as
\begin{align}\label{Inverse-corr}
    &\int_{D_A}{\rm d}^d\xi \,G^{-1}_{A,\alpha\beta}(x,\xi) G_{\beta\gamma}(\xi,\chi) = \delta_{\alpha \gamma}\delta^d(x-\chi)\,,\nonumber \\
    &\int_{D_{\bar{A}}}{\rm d}^d\chi \,G_{\alpha\beta}(\zeta,\chi) G^{-1}_{A,\beta\gamma}(\chi,\xi) = \delta_{\alpha \gamma}\delta^d(\zeta-\xi)\,,
\end{align}
and similarly for $B$. In \eqref{Inverse-corr}, $\{x, \chi\}\in D_{\bar A}$ while $\{\zeta, \xi\}\in D_{A}$.
We claim that such correlators exist in any CFT \cite{Agon:2024xvs}. Once we solve for the kernel, we plug the ansatz \eqref{eq:bilocaldef} into \eqref{mutual-twists} and obtain a formula for the MI in terms of a discrete sum over modular parameters. Such a sum can be  transformed into an integral, via   
\begin{align}
    &i\tau_{lk}\rightarrow s+\frac{i}{2}, & \lim_{n\rightarrow1} \sum_{l>k}\ \rightarrow \frac{1}{4} \int_{-\infty}^\infty \frac{\pi {\rm d}s}{\cosh^2 \pi s}\, .
\end{align}
This provides the appropriate analytic continuation, as discussed in \cite{Agon:2015ftl,Casini:2021raa,Agon:2024xvs}. 
The result reads 
\begin{widetext}
\begin{align}
    &I_\Delta=\int_{-\infty}^\infty \frac{\pi {\rm d} s}{4\cosh^2 \pi s} \int_{D_A} {\rm d}^d \xi \int_{D_B} {\rm d}^d \chi \,\Bigg|\frac{\partial \chi^J_A[s]}{\partial \chi}\Bigg|^\Delta\Bigg|\frac{\partial \xi^J_B[-s]}{\partial \xi}\Bigg|^\Delta\,\delta^d(\chi-\xi_B^J[-s])\delta^d(\xi-\chi_A^J[s])\, \nonumber\\
    &\qquad \qquad \qquad \qquad\qquad \qquad \qquad \qquad\qquad \qquad \qquad \qquad\qquad \times \text{tr}\left[ {\mathcal R}(\Lambda_A(\chi^J,s)\cdot \Lambda_B(\xi^J,-s))\right]\,, \label{eq:JacsMI}
\end{align}
\end{widetext}
where $J$ represents the action by the Tomita operator and $\chi^J_A[s]\equiv x_A[\chi,s+\frac i2]$.
The above formula can be further simplified by carrying out the remaining integrals using the delta functions. The final result is thus given in terms of a single integral over the modular parameter $s$, as shown in Eq.\,\eqref{mutual-bi-local-6} below. This represents the two-copy contribution to the MI associated to a primary operator in an arbitrary representation of the conformal group.

{\bf Relatively boosted spheres: }
The most general version of $I_{\Delta}$ is obtained by considering two disjoint spheres with a general relative boost. Without loss of generality, we consider a geometric configuration in which the spheres of radii $R_A,R_B$ are concentric and satisfy
$R_A>R_B \exp|\beta|$, where $\beta$ represents the relative boost which we apply to the smaller one. This guarantees that the smaller sphere is completely contained within the larger one. Notice that the entangling region associated to the larger sphere is the complement of a ball of radius $R_A$. This means, in particular, that its associated modular flow is $x_A[\chi,-s]$.    The flow associated to the boosted sphere can be obtained from the unboosted one via the following composition, $x_{B_\beta}[\xi,s] =\Lambda_\beta\circ x_B[\Lambda_{-\beta}(\xi),s]$, where $\Lambda_\beta$ is the boost with rapidity $\beta$ and $B_\beta$ is the associated boosted sphere of radius $R_B$. Carrying out the remaining integrals in \eqref{eq:JacsMI} leads to 
\begin{widetext}
\begin{eqnarray}\label{mutual-bi-local-6}
I_\Delta=\frac{1}{4}\int_{-\infty}^\infty \frac{\pi \,{\rm d} s}{\cosh^2\pi s} \left|\frac{\partial \chi^J_A(s)}{\partial \chi}\right|_{\chi=\chi^*}^{d-\Delta} 
 \left|\frac{\partial \chi^J_{B_\beta}(-s)}{\partial \chi}\right|^{\Delta}_{\chi=\chi^*} \, \frac{\text{tr}\left[ {\mathcal R}( \Lambda_{B_\beta}(\chi^J,-s) \cdot \Lambda^{-1}_A(\chi^J,s)) \right]_{\chi=\chi^*}}{\left|\det 
 \left( \frac{\partial f^\alpha(\chi)}{\partial \chi^\beta}\right)
 \right|_{\chi=\chi*}}\,. 
\end{eqnarray}
\end{widetext}
Here $\chi^*$ is determined by the solution to the equation $f(\chi^*)=0$ subject to the condition that $\chi^*\in D_{ A}$, where $f(\chi)\equiv x_A^J[\chi,s]-x_{B_\beta}^J[\chi,-s]$.
The solutions to this equation come in four branches, but the condition  $\chi^*\in D_{ A}$ singles out the following two,
\begin{equation}
\frac{\chi_*^\pm}{R_A} =\pm\frac{(x_\beta^\pm-1 )\cosh \pi s -(1+x_\beta^\pm)\sqrt{\cosh^2 \pi s-z^\pm}}{2\sinh \pi s}\, ,
\end{equation}
where we define $x\equiv R_B/R_A$ and their boosted analogues, namely $x_\beta^\pm \equiv e^{\pm\beta}x$. We also introduce the notation $(z^-,z^+)\equiv (z,\bar{z})$ for the conformal ratios, which in this concentric configuration read $z^{\pm} =4x_\beta^\pm/(1+x_\beta^\pm)^2$. 
Each term in \eqref{mutual-bi-local-6} can be computed explicitly---see the supplementary material.   A crucial ingredient of \eqref{mutual-bi-local-6} is the character of the representation evaluated for a particular Lorentz transformation. This corresponds to a pure boost, $\tilde{\beta}$, which depends on the modular parameter  $s$, the original boost $\beta$ and the geometric configuration $x$. Indeed, we may write
\bea \label{CharacterGeneral}
\bm{\chi}_{_\mathcal R}(\Lambda(\tilde{\beta}))\equiv \text{tr}\left[ {\mathcal R}( \Lambda_{B_\beta}(\chi^{*J},-s) \cdot \Lambda^{-1}_A(\chi^{*J},s)) \right]\,,\nonumber \\
\eea
where $\tilde{\beta}$ satisfies
\bea\label{boost-modular}
\cosh \frac{\tilde{\beta}}{2}=\frac{1-\sqrt{(1-w\,z)(1-w\bar{z})}}{w\sqrt{z \Bar{z}}}\,.
\eea 
All this leads to the main result of our paper,
\begin{widetext}
\begin{align}
    I_\Delta=\frac{z^\Delta \Bar{z}^\Delta}{2^{d+4}} \int_0^1 {\rm d}w  \frac{w^{2\Delta}}{\sqrt{1-w}} \frac{(1+\sqrt{1-w z})^{d-2\Delta}}{\sqrt{1-w z}}\frac{(1+\sqrt{1-w\Bar{z}})^{d-2\Delta}}{\sqrt{1-w\Bar{z}}} \frac{\bm{\chi}_{_\mathcal R}(\Lambda(\tilde{\beta}))}{(\sqrt{1-w z}+\sqrt{1-w\Bar{z}})^{d-2}} \,,\label{MIExplicit}
\end{align}
\end{widetext}
where the integral over $w$ is a reparametrization of the modular integral via $w\to \frac{1}{\cosh^2 \pi s}$. 

Explicit formulas for the character  of the transformation $\bm{\chi}_{\mathcal R}(\Lambda(\tilde{\beta}))$ were worked out recently in the context of MI in  \cite{Casini:2021raa} following \cite{Hirai:1965}. We give a short summary in the supplementary material, along with a few relevant examples.

In the limit where the regions are very far away, $\cosh{\tilde{\beta}}/2\approx \frac{z+\bar{z}}{2\sqrt{z\bar{z}}}$. This can be approximated in terms of normal vectors characterizing the geometry by $\cosh{\tilde{\beta}}/2\approx \cosh{\beta}= 2 (n_A\cdot l)(n_B\cdot l)-n_A\cdot n_B$, where $n_A$ and $n_B$ give the orientation of the spheres and $l$ is the normal vector that connects the past causal tips of the regions---see the supplementary material.

Another limit that is worth mentioning is the one of parallel spheres. In that case $\cosh{\tilde{\beta}}/2=0$. Therefore, $\Lambda(\tilde{\beta})=\mathbf{1}$  and $\bm{\chi}_{_\mathcal R}(\mathbf{1})= {\rm dim}\,\mathcal R $. This limit provides a simple extension of the result presented in \cite{Agon:2024xvs} to the case of arbitrary primary operators, namely
\begin{align}
    I_\Delta = {\rm dim} {\cal R} 
     \int_0^1   \frac{{\rm d}w\, (z w)^{2\Delta} \(1+\sqrt{1-wz}\)^{2(d-2\Delta)}}{4^{d+1}\sqrt{1-w}(1-wz)^{\frac{d}2}} \,. \label{MI-parallel}
\end{align}

{\bf Large separations and conformal block expansions:}
 When the regions are very far from each other, the R\'enyi twist operators become effectively local, and the MI is dominated by correlations between arbitrary pairs of points, one in each region.
 The leading contribution comes from two copies of the primary with the lowest conformal dimension, $\mathcal{O}\mathcal{O}$ \cite{Casini:2021raa}. 
Subleading contributions arising from descendants such as $\partial\mathcal{O}\mathcal{O}$, from multiple copies, as well as from other primaries become relevant as the regions approach.

 Interestingly, for ball regions, considering  the two-point functions at coincident points in the replica manifold---corresponding to the $x_2\rightarrow x_1$ limit of the general constraint (\ref{eq:two-point-twist})---completely fixes the form of the twist operator to be the OPE block: $\mathcal{O}\mathcal{O}+ \text{descendants}$. This full tower can be resumed under the integral representation $\tilde{\Sigma}_A^{(n)}\approx \int_{D(x_A^+,x_A^-)} g(\xi) \mathcal{O}(\xi)\mathcal{O}(\xi)$, where the leading primary is smeared over the causal completion of the sphere $D(x_A^+,x_A^-)$. In turn, $g(\xi)$ is a local kernel that represents the fusion of $\mathcal{O}(x_A^+)$ and $\mathcal{O}(x_A^-)$ in the channel of the shadow operator \cite{Czech:2016xec, Simmons-Duffin:2012juh}. Consequently, the resulting MI is the conformal block associated to the leading primary,  $I_{\Delta}^{\text{local}}\propto G_{2\Delta}^{\ell}(z,\bar{z})$, with $\ell$ the corresponding spin representation.

On the other hand, as explained above, the expectation value of the two copy-sector at arbitrary insertion points $x_1\neq x_2$ leads to a bilocal kernel ansatz for the twist operator---see Eq.\,\ref{eq:bilocaldef}. Following the analysis of \cite{Agon:2024xvs}, the correlator of this twist gives rise to a sum of conformal blocks, $I^{\text{bi-local}}_{\Delta}\propto \sum_i b_iG^{\ell_i}_{\tilde \Delta_i}(z,\bar z)$. Each block is associated to a different conformal primary in replica space---with conformal dimension $\tilde \Delta_i$ and spin $\ell_i$---made out of the product of two seed CFT operators, $\mathcal{O}\mathcal{O}$ being the one with lowest conformal dimension---namely, $\tilde \Delta_1=2\Delta$. This is explained in more detail in the supplementary material, where we explicitly verify the agreement between the long-distance expansion of our general formula and the corresponding conformal blocks sum in the case of the scalar representation.
In Fig. \ref{fig:Mutual} we compare the local and bilocal ansatze with the known exact result \cite{Arias:2018tmw} for a two-dimensional free chiral scalar. As expected on general grounds, the latter provides a better approximation at long distances. 


{\bf A high-precision ansatz for arbitrary separations:}
An obvious limitation of (\ref{MIExplicit}) is that it lacks information about the multiple-copy sector. Hence, although it should approximate extremely well the exact MI at long distances for general theories, it will also deviate very significantly from the expected result at short distances. In fact, for parallel, concentric balls of radii $R$ and $R+\delta$, Eq. (\ref{MI-parallel}) behaves, in $d>2$, as
\begin{align}
    \frac{I_\Delta}{{\rm dim}\mathcal{R}} &= \frac{\sqrt{\pi} \Gamma\left(\frac{d-1}{2}\right)}{2^{d+3}\Gamma\left(\frac{d}{2}\right)}\left(\frac{R}{\delta}\right)^{d-1}\label{vol_law}\\ \notag
   & -\left(\Delta-\frac{d}{2}\right)\frac{\sqrt{\pi} \Gamma\left(\frac{d-2}{2}\right)}{2^{d+2}\Gamma\left(\frac{d-1}{2}\right)}\left(\frac{R}{\delta}\right)^{d-2}+\ldots\, , 
\end{align}
when $\delta\ll R\nonumber$. 
This is a volume law divergence which severely differs from the short-distance mutual information of an ordinary local CFT
\begin{equation}
I= k_d \frac{2\pi^{\frac{d-1}{2}}}{\Gamma[\frac{d-1}{2}]}\left(\frac{R}{\delta}\right)^{d-2}+ c_{d-4}\left(\frac{R}{\delta}\right)^{d-4}+\ldots\, ,
\end{equation}
where $k_d$ is the universal coefficient appearing in the EE of a strip region \cite{Casini:2007dk,Casini:2005zv,Casini:2015woa}. Moreover, note that while (\ref{vol_law}) includes all integer powers of the expansion parameter, the exact result exclusively includes either odd or even powers of $(R/\delta)$. 

In order to find a precise approximation to the exact MI of spherical regions in an arbitrary CFT, we propose a two-fold modification of (\ref{MI-parallel}). Firstly, we change the number of spacetime dimensions from $d$ to $d-1$. Crucially, the first few leading terms in the MI long-distance expansion---see Eqs. (\ref{I_order1})---do not depend on $d$, but only on $\Delta$. Hence, this apparently drastic modification does not spoil the approximation to the exact result at long distances. Secondly, we add the contribution of an additional operator of dimension $\Delta^*$, with $\sigma$ degrees of freedom, namely,
\begin{equation}\label{MI_short_distance_ansatz}
I^{\text{Ansatz}}\equiv I_{\Delta}(d-1)+\sigma\cdot  I_{\Delta^*}(d-1)\, .
\end{equation}
In order to exactly match the area term, we fix $\sigma$ in terms of the strip coefficient, 
\begin{equation}\label{sigm}
\sigma\equiv k_d\frac{2^{d+3}\pi^{\frac{d-2}{2}}}{{\rm dim}\mathcal{R}~\Gamma[\frac{d-2}{2}]}-1\, ,   \quad (d>2)
\end{equation}
This works because, crucially again, the leading term in the short-distance expansion of $I_{\Delta}$ does not depend on $\Delta$ explicitly.
On the other hand, different strategies can be considered in order to optimize $\Delta^*$: i) we can fix it by requiring that the spurious $(R/\delta)^{d-3}$ term vanishes, which entails setting
$\Delta^*=\left(1+\frac{1}{\sigma} \right)\left(\frac{d-1}{2}\right)-\frac{\Delta}{\sigma}$,
as long as $\Delta^*>\Delta$ (so we do not spoil the long-distance behavior); ii) we can also fix it by choosing it to be equal to the second lowest primary conformal dimension of the theory. In Fig.\,\ref{fig:Mutual} we try these approaches in the case of a  $d=2$ free fermion, a $d=3$ free scalar---finding an excellent agreement with the available exact and numerical lattice results, respectively---
and a $d=4$ Maxwell field, for which our ansatz is expected to provide the best available approximation to the unknown exact result. As recently shown in \cite{Abate:2025ywp}, the $d=4$ Maxwell MI is double the difference between the $d = 4$ free scalar and $d = 2$ chiral scalar results. We can therefore alternatively approximate the Maxwell MI indirectly by first applying our method to the $d = 4$ free scalar.
We compare both approaches in Fig.\,\ref{fig:Mutual}---see the black and dashed blue curves---finding excellent agreement. 

In sum, from the knowledge of the lowest scaling dimension of a given CFT, $\Delta$, its spin representation, and its corresponding strip coefficient, $k_d$, we can construct high-precision approximations to the MI of pairs of spherical regions at arbitrary separations using (\ref{MI_short_distance_ansatz}). 

{\bf Conclusions.} In this letter we have computed the general contribution of a given primary to the MI of two balls resulting from the two-copy sector of a general $d$-dimensional CFT. Our result is most naturally applied to the operator with the lowest scaling dimension of the theory, providing the state-of-the-art long-distance approximation to the MI of any CFT. On the other hand, such formula leads to a pathological volume-law behavior in the short-distance regime. Interestingly, exploiting the dimension-independent form of the leading long-distance term and the $\Delta$-independent form of the leading short-distance one, we have put forward a new ansatz which interpolates between the exact results in both regimes. This approximates very accurately the available exact free-field  results for arbitrary separations.   

The fact that using only partial information---the two-copy sector---leads to a volume law at short distances is reminiscent of generalized free fields (GFFs), for which the same phenomenon arises \cite{Benedetti:2022aiw}.  In fact, for unboosted spheres our formula \eqref{MI-parallel} provides an excellent approximation to the MI of a GFF in an arbitrary spin represention in the full range of cross ratios---which generalizes the spin-$0$ result of \cite{Agon:2024xvs}. 
This match suggests that our formula may be capturing the contribution from some kind of coarse grained set of degrees of freedom. 

Additional future directions should also include incorporating the effects of operators from
multiple copies. Such developments would move us even closer to a full systematic characterization of any CFT in terms of vacuum MIs.

\vspace{0.1cm}
{\bf Acknowledgments:} We would like to thank Horacio Casini and Guim Planella for useful discussions.
CA is funded by the program ``Saavedra Fajardo" 22824/SF/24 from Fundaci\'on S\'eneca de la regi\'on de Murcia, and by the Spanish MINECO grant PID2024-155685NB-C22.
PB and GvdV were supported by a Proyecto de Consolidaci\'on Investigadora (CNS 2023-143822) from Spain Ministry of Science, Innovation and Universities, and by the grant PID2022-136224NB-C22, funded by MCIN/AEI/ 10.13039/501100011033/FEDER, UE. This Letter is dedicated to the memory of Umut G\"ursoy, with whom two of the authors had many fruitful discussions on the main topic of this work.
\onecolumngrid  \vspace{1cm} 
\begin{center}  
{\Large\bf Supplementary material} 
\end{center} 

\section{Derivation of the main formula}\label{QTGs2}
In the main text we have derived the intermediate formula
\begin{eqnarray}\label{mutual-bi-local-Supp1}
I_\Delta=\frac{1}{4}\int_{-\infty}^\infty \frac{\pi \,{\rm d} s}{\cosh^2\pi s} \left|\frac{\partial \chi^J_A(s)}{\partial \chi}\right|_{\chi=\chi^*}^{d-\Delta} 
 \left|\frac{\partial \chi^J_{B_\beta}(-s)}{\partial \chi}\right|^{\Delta}_{\chi=\chi^*} \, \frac{\text{tr}\left[ {\mathcal R}( \Lambda_{B_\beta}(\chi^J,-s) \cdot \Lambda^{-1}_A(\chi^J,s)) \right]_{\chi=\chi^*}}{\left|\det 
 \left( \frac{\partial f^\alpha(\chi)}{\partial \chi^\beta}\right)
 \right|_{\chi=\chi*}}\,.
\end{eqnarray}
In this appendix we compute the various terms appearing above and we  explain how to derive our  final expression for the contribution to the MI of two relatively boosted spheres arising from the two-copy sector associated to a primary of conformal dimension $\Delta$---see \eqref{MIExplicit_Supp} below.

First, we must evaluate the Jacobian of the transformations $\chi^J_A(s)=x_A[\chi,s+i/2]$ and $\xi^J_{B_\beta}(-s)=x_{B_\beta}[\xi,-s+i/2]$ which are defined as
\bea\label{Jacobian}
\left|\frac{\partial x^J_A(s)}{\partial x}\right|=\left|{\rm det}\left(\frac{\partial x^{\alpha }_A(s+i/2)}{\partial x^\beta} \right)\right|^{1/d}\,, \quad {\rm where}\quad \frac{\partial x^{\alpha }_A(s+i/2)}{\partial x^\beta} = \left|\frac{\partial x^J_A(s)}{\partial x}\right| \left(\Lambda_A\right)^{\alpha}_{\,\,\beta}(x^J,s)\,.
\eea
The formulas are identical for $\xi_{B_\beta}^J(s)$ but with the modular flow of $B_\beta$ instead of $A$. The determinant of the modular flow of a sphere is known to be:
\begin{align}
    &\left|\frac{\partial x^J_A(s)}{\partial x} \right| = \frac{R_A^2\, \text{sech}^2 \pi s }{(R_A\tanh \pi s+x^+)(R_A\tanh\pi s-x^-)} \,,\label{det_MF}
\end{align}
where $x^\pm=r\pm t$ are null coordinates. From here, computing the Jacobian of $\chi_A^J(s)$ simply entails inserting the point $\chi_*$ which solves $f(\chi_*)=0$, with $f(\chi)\equiv x_A^J[\chi,s]-x_{B_\beta}^J[\chi,-s]$, into the formula \eqref{det_MF} above. The exact form of $\chi_*$ can be found in the main text.  For $\xi_{B_\beta}^J(-s)$ the procedure is similar. Recall that $x_{B_\beta}[\xi,s]=\Lambda_\beta (x_{B}[\Lambda_{-\beta}(\xi),s])$, where $\Lambda_\beta(\cdot)$ is a boost with rapidity $\beta$ and $x_B[\xi,s]$ is the modular flow of the unboosted sphere. For the Jacobian this means that we have to evaluate the determinant of the modular flow of a sphere of radius $R_B$ at the point $\Lambda_{-\beta} (\chi_*)$ instead. After some algebra, the two determinants can be evaluated and read
\begin{align}\label{detJA}
\left|\frac{\partial \chi^J_A(s)}{\partial \chi}\right|_{\chi=\chi^*} &=\frac{\(1+\sqrt{1-w \,z}\)\(1+\sqrt{1-w \,\bar{z}}\)}{z \bar{z} \,w (1-w)}\(\frac{1-x_\beta^-}{1+x_\beta^-} -\sqrt{1-w \,z}\)\(\frac{1-x_\beta^+}{1+x_\beta^+} -\sqrt{1-w \,\bar{z}}\)\,,\\  \label{detJB}
 \left|\frac{\partial \chi^J_{B_\beta}(-s)}{\partial \chi}\right|_{\chi=\chi^*} &=\frac{w}{1-w}\frac{\(\frac{1-x_\beta^-}{1+x_\beta^-} -\sqrt{1-w \,z}\)\(\frac{1-x_\beta^+}{1+x_\beta^+} -\sqrt{1-w \,\bar{z}}\)}{\(1+\sqrt{1-w \,z}\)\(1+\sqrt{1-w \,\bar{z}}\)}\,,
\end{align}
where we have introduced the reparameterization $w=1/\cosh^2\pi s$, and $x_\beta^\pm=e^{\pm \beta}x$, $z^\pm=4 x_\beta^\pm/(1+x_\beta^\pm)^2$, with $(z^-,z^+)\equiv (z,\bar{z})$.
We also need the determinant of the transformation $\partial f^\alpha/\partial \chi^\beta$ for $f(\chi)$ given above. For that purpose, it is convenient to separate the Jacobians from the  modular transformations according to \eqref{Jacobian}. Specifically, we can  re-write the Jacobian as $\frac{\partial f}{\partial \chi} =\frac{\partial f}{\partial \chi} \cdot \Lambda^{-1}_A(\chi^J,s)\cdot \Lambda_A(\chi^J,s)$. The determinant of this matrix can be decomposed as follows,
\bea
\det\(\frac{\partial f^\alpha(\chi)}{\partial \chi^\beta}\)=\det \(\left|\frac{\partial \chi^J_A(s)}{\partial \chi}\right| {\mathbf{1}}-\left|\frac{\partial \chi^J_{B_\beta}(-s)}{\partial \chi}\right|\(\Lambda_{B_\beta}(\chi^{J},-s) \cdot \Lambda^{-1}_A(\chi^{J},s))\)\) 
\det\( \Lambda_{A}(\chi^J,s)\)\,.
\eea
Moreover, the composed transformation $\Lambda_{B_\beta}(\chi^{*J},-s) \cdot \Lambda^{-1}_A(\chi^{*J},s))$ corresponds to a boost $\tilde{\beta}$ given by 
\bea\label{boost-modular-Supp}
\cosh \frac{\tilde{\beta}}{2}=\frac{1-\sqrt{(1-w\,z)(1-w\bar{z})}}{w\sqrt{z \Bar{z}}}\,.
\eea 
Therefore, in its diagonal basis, it has the form diag$\left\{e^{\tilde \beta},e^{-\tilde{\beta}}, {\mathbf 1}_{d-2}\right\}$. Computing the determinant is now nothing more than multiplying the diagonal elements. This leads to
\bea\label{detf-chi-2}
\left|\det\(\frac{\partial f^\alpha(\chi)}{\partial \chi^\beta}\)\right|=\left|\left|\frac{\partial \chi^J_A(s)}{\partial \chi}\right|- \left|\frac{\partial \chi^J_{B_\beta}(-s)}{\partial \chi}\right| \right|^{d-2}\left|\left|\frac{\partial \chi^J_A(s)}{\partial \chi}\right|- \left|\frac{\partial \chi^J_{B_\beta}(-s)}{\partial \chi}\right|e^{\tilde{\beta}} \right|\left|\left|\frac{\partial \chi^J_A(s)}{\partial \chi}\right|- \left|\frac{\partial \chi^J_{B_\beta}(-s)}{\partial \chi}e^{-\tilde{\beta}}\right| \right| \,,\nonumber\\
\eea
where we use the fact that $\left|\det\( \Lambda_{A}(\chi^J,s)\)\right|=1$.
Plugging the formula for $\Tilde{\beta}$, \eqref{boost-modular-Supp}, and the expressions for the Jacobians, \eqref{detJA} and \eqref{detJB}, into \eqref{detf-chi-2} leads to 
\bea
\left|\det\(\frac{\partial f^\alpha(\chi)}{\partial \chi^\beta}\)\right|_{\chi=\chi^*}\!\!\!\!=\frac{2^{d+2}\sqrt{1-w \,z}\sqrt{1-w \,\bar{z}}}
{ \,z^d \,\bar{z}^d \,w^d\, (1-w)^d}
\frac{\(\frac{1-x_\beta^-}{1+x_\beta^-} -\sqrt{1-w \,z}\)^d\(\frac{1-x_\beta^+}{1+x_\beta^+} -\sqrt{1-w \,\bar{z}}\)^d}
{\(\sqrt{1-w \,z} +\sqrt{1-w \,\bar{z}}\)^{2-d}}\,,
\eea
where once again we wrote the answer in the parametrization $w=1/\cosh^2\pi s$. Putting all the Jacobians together in \eqref{mutual-bi-local-Supp1} we obtain the final expression 
\begin{align}
    I_\Delta=\frac{z^\Delta \Bar{z}^\Delta}{2^{d+4}} \int_0^1 {\rm d}w  \frac{w^{2\Delta}}{\sqrt{1-w}} \frac{(1+\sqrt{1-w z})^{d-2\Delta}}{\sqrt{1-w z}}\frac{(1+\sqrt{1-w\Bar{z}})^{d-2\Delta}}{\sqrt{1-w\Bar{z}}} \frac{\bm{\chi}_{_\mathcal R}(\Lambda(\tilde{\beta}))}{(\sqrt{1-w z}+\sqrt{1-w\Bar{z}})^{d-2}} \,.\label{MIExplicit_Supp}
\end{align}

\section{Computing the Character and Various Examples}
The dependence of MI on the spin is completely captured by the character ($\chi_\mathcal{R}(\Lambda(\Tilde{\beta}))$) in the integral \eqref{MIExplicit_Supp}. The general form of the  character for a representation $\mathcal{R}$  is given by
\bea \label{CharacterGeneral-Supp}
\bm{\chi}_{_\mathcal R}(\Lambda(\tilde{\beta}))\equiv \text{tr}\left[ {\mathcal R}( \Lambda_{B_\beta}(\chi^{*J},-s) \cdot \Lambda^{-1}_A(\chi^{*J},s)) \right]\, .
\eea
The Lorentz transformations in this formula look complicated. However, the character only depends on the eigenvalues of the Lorentz transformation $\Lambda_{B_\beta}(\chi^{*J},-s)\cdot \Lambda_A^{-1}(\chi^{*J},s)$ which are given by $(e^{\Tilde{\beta}},e^{-\Tilde{\beta}},1,\dots,1)$, where the boost parameter $\Tilde{\beta}$ satisfies \eqref{boost-modular-Supp}.

The character of an arbitrary representation of the Lorentz group was studied in \cite{Hirai:1965}. However, direct applications of these formulas for a Lorentz transformation with the spectrum $(e^{\Tilde{\beta}},e^{-\Tilde{\beta}},1,\dots,1)$, give indeterminate expressions of the form 0/0. One has to consider instead eigenvalues of the form $(e^{\Tilde{\beta}},e^{-\Tilde{\beta}},e^{i\epsilon_1},e^{-i\epsilon_1},\dots ,e^{i\epsilon_q},e^{-i\epsilon_q},1)$ for odd $d$ and $(e^{\Tilde{\beta}},e^{-\Tilde{\beta}},e^{i\epsilon_1},e^{-i\epsilon_1},\dots ,e^{i\epsilon_q},e^{-i\epsilon_q})$ for even $d$. After computing the character, one takes the limit $\epsilon_i\rightarrow 0$. This has already been studied in the Appendix B of \cite{Casini:2021raa}, so we will not repeat the computation here. Instead, we will simply collect the results.

A Lorentz representation is determined by the lengths of the rows of its Young diagram. Let the representation $\mathcal{R}$ have rows of length $m_1\leq m_2 \leq \dots \leq m_q$, where $q$ is $q=\lfloor \frac{d}{2}-1 \rfloor$. The characters for odd and even dimensions are given below: 
\begin{align}
    &\chi_\mathcal{R}(\Lambda(\Tilde{\beta})) = \frac{\sum_{j=1}^{q+1}(-1)^{j+q+1}\sinh(\Tilde{\beta}(m_j+j-\frac{1}{2}))\det (A_j)}{\sinh(\Tilde{\beta}/2)(\cosh \Tilde{\beta}-1)^q} & (d \text{ odd)}
    \, ,\\
    &\chi_\mathcal{R}(\Lambda(\Tilde{\beta})) = \frac{\sum_{j=1}^{q+1}(-1)^{j+q+1}\cosh(\Tilde{\beta}(m_j+j-1))\det (B_j)}{(\cosh \Tilde{\beta}-1)^q} & (d \text{ even)\,,}
\end{align}
where the matrices $A_j$ and $B_j$ are defined as
\begin{align}
    &A_j = \Big\{\frac{(m_k+k-\frac{1}{2})^{2l-1}}{(2l-1)!} \Big\}^{k=1,\dots, \Hat{j},\dots,q+1}_{l=1,\dots,q}\, ,\\
    &B_j=\Big\{\frac{(m_k+k-1)^{2l}}{(2l)!} \Big\}^{k=1,\dots, \Hat{j},\dots,q+1}_{l=0,\dots,q-1  }\, .
\end{align}
Here, $\Hat{j}$ means that the $j$'th element is omitted. It is easy to check that in the absence of boost, $\beta=0$, \eqref{boost-modular-Supp} reduces to $\Tilde{\beta}=0$, giving us the identity element. The character of the identity is the dimension of the representation $\dim \mathcal{R}$. 

Below we list a number of examples for various representations. For a spin-$\frac{1}{2}$ primary we have:
\begin{align}
    &\chi_{\frac{1}{2}}(\Lambda(\Tilde{\beta})) = \frac{1-\sqrt{1-wz}\sqrt{1-w\Bar{z}}}{w\sqrt{z\Bar{z}}}\, .
\end{align}
For a spin-1 (vector $A_\mu$) we just have a Young diagram with one box $m_1=1$ or, alternatively, one can note that the character of spin-1 is just the sum of the eigenvalues, $2\cosh \Tilde{\beta}+d-2$. The character reads
\begin{align}
    &\chi_{1}(\Lambda(\Tilde{\beta})) = 4\left(\frac{1-\sqrt{1-wz}\sqrt{1-w\Bar{z}}}{w\sqrt{z\Bar{z}}}\right)^2+d-4\, .
\end{align}
Lastly rank-2 anti-symmetric tensors $F_{\mu\nu}$ have a Young diagram with two vertical boxes, hence $m_1=m_2=1$. After some computations the character reads
\begin{align}
    &\chi_{2}(\Lambda(\Tilde{\beta})) = 4(d-2)\left(\frac{1-\sqrt{1-wz}\sqrt{1-w\Bar{z}}}{w\sqrt{z\Bar{z}}}\right)^2+\frac{(d-4)^2-d}{2}\, .
\end{align}

\section{Long Distance Limit}\label{long_distance_limit}
In this appendix we compute the long-distance expansion of the MI for scalar CFTs, which corresponds to taking $\bm{\chi}_{_\mathcal R}(\Lambda(\tilde{\beta}))=1$ in (\ref{MIExplicit_Supp}). More explicitly, we take the limit in which the non-parallel spheres are very far from each other. To accomplish this, it is convenient to work with the parameters $\chi_1$ and $\chi_2$, related to the cross ratios $u$  and $v$ through
\begin{equation}\label{uvschi}
u=\chi_1^2\, , \quad v=\frac{\chi_1^2}{\chi_2^2}\, .
\end{equation}
In terms of the radii of the spheres $R_A$, $R_B$, the timelike unit vectors $n_A$, $n_B$ and the vector joining the lower tips of the causal regions $L l$, these parameters read \cite{Agon:2021zvp}
\begin{equation}
\chi_1=\frac{4 R_A R_B}{L\vert 2 R_A n_A-Ll-2R_Bn_B\vert}\, ,\quad \chi_2=\frac{4 R_A R_B}{\vert 2 R_A n_A-Ll\vert \vert 2R_Bn_B+Ll\vert}\, .
\end{equation}
In the limit $L\gg R_A, R_B$ \cite{Casini:2021raa},
\begin{eqnarray}\nonumber
\chi_1&=&\frac{4 R_A R_B}{L^2}+\frac{8 R_A R_B}{L^3}\left[l\cdot(R_A n_A-R_B n_B)\right]+\\
&~&\frac{8 R_A R_B}{L^4}\left[R_A^2+R_B^2+3 (l\cdot(R_A n_A-R_B n_B))^2+2R_AR_B (n_A\cdot n_B)\right]+\mathcal{O}(L^{-5})\\\nonumber
\chi_2&=&\frac{4 R_A R_B}{L^2}+\frac{8 R_A R_B}{L^3}\left[l\cdot(R_A n_A-R_B n_B)\right]+\\
&~&\frac{8 R_A R_B}{L^4}\left[R_A^2+R_B^2+3 (l\cdot(R_A n_A-R_B n_B))^2+4R_AR_B (l\cdot n_A)(l\cdot n_B)\right]+\mathcal{O}(L^{-5})\, .
\end{eqnarray}
Without loss of generality we impose $R_A=R_B$ to simplify the expressions. Furthermore, using (\ref{uvschi}) and 
\begin{equation}
z\bar{z}=u \, ,\quad (1-z)(1-\bar{z})=v\, ,
\end{equation}
we get
\begin{eqnarray}\label{zpm}
z^\pm&=&\frac{4 R^2}{L^2}e^{\pm \beta}+\frac{8 R^3}{L^3}\left(l\cdot(n_A-n_B)\right)\frac{e^{\pm 3\beta/2}}{\cosh{(\beta/2)}}+\mathcal{O}(L^{-4})\, ,
\end{eqnarray}
where for notational convenience we used $(z,\bar{z})\equiv(z^-,z^+)$ and $\cosh{\beta}\equiv 2(l\cdot n_A)(l\cdot n_B)-(n_A\cdot n_B)$. Note that this consistently reduces to $z=\bar{z}=4R^2/L^2$ in the parallel limit ($\beta\rightarrow 0$). This also implies that we can expand (\ref{MIExplicit_Supp}) for small $z$. The first three terms in the small $z$ expansion are
\begin{eqnarray}\nonumber
I_{(1)}&=&\frac{\sqrt{\pi}\Gamma[1+2\Delta]}{ 16^\Delta4\Gamma[\frac{3}{2}+2\Delta]}(z\bar{z})^\Delta\, ,\\  \nonumber 
I_{(2)}&=&\frac{\sqrt{\pi}\Delta\Gamma[2+2\Delta]}{ 16^\Delta8\Gamma[\frac{5}{2}+2\Delta]}(z\bar{z})^\Delta(z+\bar{z})\, ,\\  \label{I_order1}
I_{(3)}&=&\frac{\sqrt{\pi}\Gamma[3+2\Delta]}{ 16^\Delta64\Gamma[\frac{7}{2}+2\Delta]}(z\bar{z})^\Delta\left((1+3\Delta+2\Delta^2)(z^2+\bar{z}^2)+(d-2+4\Delta^2)z\bar{z}\right)\, .
\end{eqnarray}
Following the analysis of \cite{Agon:2024xvs}, we write the MI as a sum of conformal blocks. We will check that the conformal families contributing to (\ref{MIExplicit_Supp}) belong to the two-copy sector of replica space. At third order in a small $z$, $\bar{z}$ expansion, the terms that we must consider are
\begin{equation}\label{conformal_block_expansion}
I=b_1 G_{2\Delta}^{\ell=0}(z,\bar{z})+b_2 G_{2\Delta+1}^{\ell=1}(z,\bar{z})+b_3 G_{2\Delta+2}^{\ell=0}(z,\bar{z})+b_4 G_{2\Delta+2}^{\ell=2}(z,\bar{z})+\ldots
\end{equation}
The subindex of each conformal block labels the conformal weight $\tilde{\Delta}$ of the generating primary, while $\ell$ stands for its spin representation. Explicitly, the replica primaries involved are \cite{Agon:2024xvs}
\begin{eqnarray} \nonumber 
\mathcal{O}_i\mathcal{O}_j\, ,& \quad \tilde{\Delta}=2\Delta\, , &\quad \ell=0\\ \nonumber 
\mathcal{O}_i\partial_{\alpha}\mathcal{O}_j-\partial_{\alpha}\mathcal{O}_i\mathcal{O}_j\, ,& \quad \tilde{\Delta}=2\Delta+1\, , &\quad \ell=1\\  \nonumber 
\mathcal{O}_i\partial^2\mathcal{O}_j+\mathcal{O}_j\partial^2\mathcal{O}_i-\frac{2\Delta-d+2}{\Delta}\partial\mathcal{O}_i\cdot \partial\mathcal{O}_j \, ,& \quad \tilde{\Delta}=2\Delta+2\, , &\quad \ell=0\\    S_{\alpha\beta}
-\frac{g_{\alpha\beta}}{2}g^{\sigma\rho}S_{\sigma\rho}\, ,& \quad \tilde{\Delta}=2\Delta+2\, , &\quad \ell=2\, ,
\end{eqnarray}
with 
\begin{equation}
S_{\alpha\beta}=\partial_\alpha\partial_\beta\mathcal{O}_i\mathcal{O}_j+\mathcal{O}_i\partial_\alpha\partial_\beta\mathcal{O}_j-\frac{2\Delta+2}{\Delta}\partial_{(\alpha}\mathcal{O}_i \partial_{\beta)}\mathcal{O}_j \, .
\end{equation}
The conformal blocks for $\ell=0$ and $\ell=1$ were computed analytically in \cite{Dolan:2000ut}, these being combinations of Appell functions. Moreover, the $\ell=1$ channel can be easily computed from the last two thanks to a recursive relation. We have
\begin{eqnarray}
G_{2\Delta}^{\ell=0}(z,\bar{z})&=& (z\bar{z})^{2\Delta}\left[1+\frac{1}{2} \Delta   (z+\bar{z})+\frac{1}{4} \Delta \left(\frac{4 \Delta ^2 z \bar{z}}{(2 \Delta +1) (-d+4 \Delta +2)}+\frac{(\Delta +1)^2 (z+\bar{z})^2}{2 \Delta +1}-2 z \bar{z}\right)+\mathcal{O}\left(z ^3\right)\right]\nonumber\\
G_{2\Delta+1}^{\ell=1}(z,\bar{z})&=& (z\bar{z})^{2\Delta}\left[-\frac{1}{2}  (z+\bar{z})-\frac{1}{4} \left(z^2+\Delta  (z+\bar{z})^2+\bar{z}^2\right)+\mathcal{O}\left(z ^3\right)\right]\nonumber\\
G_{2\Delta+2}^{\ell=0}(z,\bar{z})&=& (z\bar{z})^{2\Delta}\left[(z\bar{z})^2+\mathcal{O}\left(z^3\right)\right]\nonumber\\
G_{2\Delta+2}^{\ell=2}(z,\bar{z})&=& (z\bar{z})^{2\Delta}\left[\frac{ \left(d (z+\bar{z})^2-4 z \bar{z}\right)}{4 d}+\mathcal{O}\left(z^3\right)\right]\, .
\end{eqnarray}
On the other hand, the coefficients in (\ref{conformal_block_expansion}) are computed by requiring that the twist operator produces modular
evolution within correlation functions. These read \cite{Agon:2024xvs}
\begin{eqnarray}\label{b1}
b_1&=&\frac{\sqrt{\pi}\Gamma(2\Delta+1)}{2^{4\Delta+2}\Gamma(2\Delta+3/2)}\, ,\nonumber\\
b_2&=&\frac{\sqrt{\pi}\Delta\Gamma(2\Delta+1)}{2^{4\Delta+3}\Gamma(2\Delta+5/2)}\, ,\nonumber\\
b_3&=&\frac{\sqrt{\pi}(d-2-2\Delta)\Gamma(2\Delta+1)}{2^{4\Delta+4}\Gamma(2\Delta+7/2)}\left(\frac{(d-2)^2+(d-2)(3d-4)\Delta+2(3+d(d-2))\Delta^2}{2 d(d-2-4\Delta)}\right)\, ,\nonumber\\
b_4&=& \frac{\sqrt{\pi }(\Delta  (3 \Delta +4)+2) \Gamma (2 \Delta +3)}{ 2^{4 \Delta +5} (2 \Delta +1)^2 \Gamma \left(2 \Delta +7/2\right)}\, .
\end{eqnarray}
Using the above expansion of the conformal blocks and the expression of the coefficients, we can straightforwardly check that (\ref{conformal_block_expansion}) exactly matches (\ref{I_order1}). 

\section{Lattice calculations for the $d=3$ free scalar}
In this appendix we provide some details about the lattice calculations we have performed for the MI of a pair of round disk regions in the case of a three-dimensional free scalar field. 
We consider the case of unboosted disks so our lattice is two-dimensional.  Let $\phi_i,   \pi_j$ be a discrete set of  scalar fields and  conjugate   momenta defined in a square lattice of $N$ points, $i,   j= 1, \dots, N$---so that each subindex $i$ refers to a particular position $(x_i,y_i)$ in the lattice. The fields satisfy the standard commutation relations, 
\begin{equation}
    [\phi_i ,\pi_j]=i\delta_{ij}\quad  \text{and}\quad [\phi_i,\phi_j] =[\pi_i, \pi_j]=0\, .
\end{equation}
Given a  Gaussian state   $\rho$, the entanglement entropy  for some region $A$ can be obtained  from the correlators
\begin{equation}
X_{ij}\equiv \tr (\rho \phi_i\phi_j)\quad  \text{and}    \quad P_{ij}\equiv \tr (\rho   \pi_i \pi_j)\, 
\end{equation}
as follows \cite{2003JPhA...36L.205P,Casini:2009sr},
\begin{equation}\label{see}
S_A=\tr \left[(C_A +1/2) \log (C_A    +1/2)- (C_A -1/2)   \log (C_A-1/2) \right]\, , 
\end{equation}
where 
\begin{equation}
C_A \equiv  \sqrt{X_A    P_A}\quad  \text{and} \quad   (X_A)_{ij}\equiv X_{ij}\, , \quad (P_A)_{ij}\equiv P_{ij}\, ,
\end{equation}
are the restrictions of the correlators   to the lattice  sites inside the region  $A$.

The MI is computed using
\begin{equation}
    I(A,B)\equiv S_A+S_B-S_{A\cup B}
\end{equation}
and (\ref{see}).

Setting the lattice spacing to one, the discretized Hamiltonian reads
\begin{equation}
H= \frac{1}{2}   \sum_{i,j=-\infty}^{\infty}   \left[\pi^2_{i,j}   + (\phi_{i+1,j}   -\phi_{i,j})^2  +(\phi_{i,j+1}-\phi_{i,j})^2 \right]\,.
\end{equation}
The relevant vacuum-state correlators are given by \cite{Casini:2009sr} 
\begin{align}
X_{(0,0),(i,j)} & = \frac{1}{8\pi^2} \int_{-\pi} ^{\pi} {\rm d}   x  \int_{-\pi}^{\pi}  {\rm d}  y \frac{\cos (j y)  \cos (ix) }{\sqrt{2(1-\cos x)+2(1-\cos y)}}\, , \\
P_{(0,0),(i,j)} & = \frac{1}{8\pi^2}\int_{-\pi} ^{\pi}  {\rm d}    x \int _{-\pi}^{\pi}  {\rm d}  y  \cos(j y)   \cos(ix) \sqrt{2(1-\cos x)+2(1-\cos y)}\, .
\end{align}
For our calculations we make use of a square lattice of size 200.
 In the continuum limit, the lattice model approximates the results corresponding to the CFT of a $d=3$ free scalar.  In order to extract such values, we compute the MI for each possible relative distance $x\equiv D/L$---where $D$ is the separation between the disks separation and $L$ the disks radii. Keeping $x$ fixed in each case, we produce series of values for increasing values of $D,L$. The continuum result, which is achieved asymptotically for $D,L\rightarrow +\infty$ is obtained in each case by performing fits with functions $\{1,1/X,1/X^2,\dots\}$ to those series.
 Reliable continuum values are obtained when the resulting constants do not depend on the order at which we stop introducing fitting functions.

\twocolumngrid 

\bibliographystyle{JHEP-2}
\bibliography{N-partite-info-v4}
\noindent 


\end{document}